\def\Murcia{Departamento de Matem\'atica Aplicada, Facultad de Inform\'atica, Campus de 
    Espinardo, 30100 Murcia, Spain} 
 
\def\CarlosI{Instituto de F\'\i sica Te\'orica y Computacional Carlos I, 
Facultad de Ciencias, Universidad de Granada, Campus de Fuentenueva, Granada 18002, 
Spain} 
 
\def\IAA{Instituto de Astrof\'{\i}sica de Andaluc\'{\i}a, Apartado Postal 3004, Granada 
       18080, Spain} 
\def\Comision{Work partially supported by the DGICYT.}

\def\l[{\left[} 
\def\r]{\right]}

\def\R{\mathbb R}

\def\nn{\nonumber} 
\def\noi{\noindent}

\def\be{\begin{equation}} 
\def\ee{\end{equation}} 
\def\bea{\begin{eqnarray}} 
\def\eea{\end{eqnarray}} 
\def\ba{\begin{array}} 
\def\ea{\end{array}}

\documentclass[12pt]{article} 

\usepackage{amsfonts,amssymb}

\begin{document} 
 
\begin{center} 
{\LARGE {\bf Gauge principle revisited: towards a unification of
space-time and internal gauge interactions$^1$}} 
\end{center} 
 
\bigskip 
\bigskip 
 
\centerline{V. Aldaya$^{2,3}$, J.L. Jaramillo$^{2,3}$ 
       and J. Guerrero$^{2,3,4}$ } 
 
\bigskip 
 
\footnotetext[1]{\Comision} 
\footnotetext[2]{\IAA} \footnotetext[3]{\CarlosI} 
\footnotetext[4]{\Murcia} 
 
\bigskip 
 
\begin{center} 
{\bf Abstract} 
\end{center} 
 
\small 
\setlength{\baselineskip}{12pt} 
 
\begin{list}{}{\setlength{\leftmargin}{3pc}\setlength{\rightmargin}{3pc}} 
\item The minimal coupling principle is revisited under the quantum perspectives
of the space-time symmetry. This revision is better realized on a Group Approach to
Quantization (GAQ) where group cohomology and extensions of groups play a
preponderant role. We firstly consider the case of the electromagnetic
potential; the Galilei and/or Poincar\'e  group is (non-centrally)
extended by the ``local" $U(1)$ group. This group can also be seen as a central
extension, parametrized by both the mass and the electric charge, of an 
infinite-dimensional group, on which GAQ leads to the dynamics
of a particle moving in the presence of an electromagnetic field. Then
we try the gravitational interaction of a particle by making   
the space-time translations ``local". However, promoting to ``local" the
space-time subgroup of the true symmetry of the quantum free relativistic particle,
i.e. the centrally extended by $U(1)$  Poincar\'e group, results in a new
electromagnetic-like force of pure gravitational origin. This is a
consequence of the space-time translations not being an invariant subgroup of the
extended Poincar\'e group and constitutes a preliminary attempt to a
non-trivial mixing of space-time and internal gauge interactions. 
\end{list} 
 
\normalsize

\noi PACS: 02.40.-k, 03.65.Fd, 11.15.-q 

\vskip 1cm 

\pagebreak
 
\section{Introduction}
 
In the Lagrangian formalism, formulated on the $1$-jet bundle $J^1(E)$
of a vector bundle $E$ on Minkowski space-time $M$, promoting a given 
underlying rigid symmetry to ``local'', i.e. 
extending the corresponding Lie algebra by taking the tensor product of
it by the algebra of real analytic functions on $M$, requires the
introduction of a derivation law on the module of sections
of $E$, $\Gamma(E)$, which is eventually interpreted as a potential providing 
the corresponding gauge interaction. This is essentially the formulation 
of the so-called Minimal Coupling Principle, which culminates in
Utiyama's theory \cite{Utiyama}. Internal gauge invariance had originally led 
successfully to electromagnetic interaction associated with $U(1)$, then to 
Yang-Mills associated with isospin $SU(2)$ (valid only at the ``very strong" limit), 
electroweak with $(SU(2)\otimes U(1))/Z_2$, and finally to strong
interaction associated with colour $SU(3)$. The same spirit is shared by later
attempts to unify all of these into gauge groups such as
$SU(5)$. On the other hand, the ``local'' invariance under external 
(space-time) symmetries, such as a subgroup of the Poincar\'e group, has been used to 
provide	a gauge framework for gravity \cite{Kibble}, although fully
disconnected from the other (internal) interactions. In fact, a
unification of gravity and the other interactions would have supposedly required
the non-trivial mixing of the space-time group and
some internal symmetry, a task explicitely forbidden by the so-called 
{\it no-go theorems} by O'Raifeartaigh, Coleman, Mandula, Michel, etc.
\cite{O'Raifeartaigh,Coleman-Mandula,Michel,Dyson}
long ago, which stated that there is no finite-dimensional Lie group
containing the Poincar\'e group acting as diffeomorphisms of the
base manifold $M$, the Minkowski space-time, and any internal $SU(n)$
group acting linearly on the fibre of $E$,  except for the
direct product. It is worth mentioning that supersymmetry was 
originally developped in the 70's, mainly by Salam and Strathdee
\cite{supersimetria}, in an unsuccessful attempt to invalidate the no-go theorems. 
 
However, the current skill in dealing with Lie group extensions and
irreducible representations of (even infinite-dimensional) Lie groups 
tempts us into revisiting the question of the mixing of symmetries and,
accordingly, the unification of interactions in terms of ordinary
Lie groups.
We propose a simple, yet non-trivial, way of facing the problem of
interaction mixing. This consists in identifying one of the
$U(1)$ Cartan subgroups in the internal symmetry with the $U(1)$ phase
invariance in Quantum Mechanics.
Then, turning the space-time translation subgroup of a centrally 
extended Poincar\'e group \cite{Saletan,pseudo,JPA,Conforme} into a
``local'' group automatically promotes the
original rigid internal symmetry to the gauge level in a non-trivial way
from the physical standpoint. This provides a non-trivial mixing of 
gravity and the already introduced internal interaction associated with the given
unitary symmetry. Here, we seek to demonstrate explicitly the ocurrence
of this new phenomenon at least at a given approximation without
exhausting all possibilities of the proposed algorithm. This means that
our present computational outputs must be understood as being partial,
although, at the non-relativistic limit, they are exact and reveal with
precision the mixing effect \cite{electrograv}.

As is wellknown, the Minimal Coupling Principle can also be formulated on a bundle $E$ of
the form $E={\R}^3\times \R\rightarrow \R$, as corresponding to the case of
Mechanics, thus making the problem technically easier while keeping the essential point to
be discussed here. In this framework, to be followed in the present paper,
the gauge principle will be revisited
by simply substituting the Quantum
Mechanical space-time symmetry for the standard (classical) one. In fact,
in a previous letter \cite{electrograv}, we sketched an approach to the problem in
this simple and economical way, i.e. in a Particle Mechanics (versus Field Theory)
framework, and we now undertake a  more detailed and formal presentation. A
much more involved generalization to Quantum Field Theory is under way \cite{infinito}. 

Since the revision we attempt here lies entirely on symmetry grounds, a
(Quantum) Mechanical formulation tightly attached to a group structure
is better suited. There is, in fact, a way of associating
physical dynamics with a specific symmetry group. This could be accomplished  
by means of the rather standard co-adjoint-orbits method of Kirillov 
\cite{Kirillov}, where the Lagrangian is seen as the local potential of
the corresponding symplectic form. However, we shall proceed through   
a group approach to quantization (GAQ) \cite{23,Ramirez} which is
directly related to the co-homological structure of the symmetry group,
and leads directly to the quantum theory, a fact that can be of capital
relevance in the near future in passing to the quantum-field-theory level. 
Co-homology parameters will be directly identified with the
physical coupling constants. In this sense, the association of the parameter of the
(symplectic) co-homology group of the Galilei group with the particle
mass has been emphasized by Souriau \cite{Souriau}.

As an intermediate step between the more standard Lagrangian version of the
Minimal Interaction Principle and the one to be presented here, we shall
formulate a version ``a la Cartan", i.e. in terms of the invariance of the
Poincar\'e-Cartan form \cite{Malliavin,Godbillon,Abraham,Marsden} rather than
the Lagrangian, of the part of Utiyama's theory
concerning the particle in interaction with the field. Indeed, as
mentioned above, the dynamics of the gauge fields themselves will be considerd elsewhere.

The paper is organized as follows. Sec. II is a thorough presentation of
basic geometrical aspects of Classical and Quantum Mechanics, mainly those 
fundamental to the development of the present work. Sec. III is
devoted to the Cartan-like analysis of the Minimal Coupling
Principle particularized for the case of electromagnetism and
non-relativistic gravity. In Sec. IV we present explicitly the GAQ
with the example of a particle moving in an electromagnetic
field. Finally, in Sec. V, we directly present the chief problem of
gauging the translation subgroup of the centrally extended Poincar\'e
group giving rise to the new phenomenon of an extra coupling constant
mixing non-trivially the geodesic force and the Lorentz one. Some
outlooks are included at the end.  

\medskip

\section{Phase invariance in Quantum Mechanics}

\medskip

According to the standard approach to Quantum Mechanics (see for instance
\cite{Roman}) a state of the system is characterized by a {\it ray},
rather than a vector, of a Hilbert space i.e. normalized wave functions
are determined up to a complex number of module $1$ or phase. This is a
direct consequence of the definition of probability and constitutes a
symmetry to be referred to as $U(1)$ or Phase Invariance in Quantum Mechanics.
Let us approach this symmetry from quite different
perspectives to highlight its fundamental features. 

\medskip
\subsection{Behaviour of the Schr\"odinger equation}
\medskip

We shall consider the behaviour of the Schr\"odinger equation
corresponding to the free quantum particle

\begin{equation}
i\hbar\frac{\partial \,}{\partial t}\Psi=-\frac{\hbar^2}{2m}\nabla^2\Psi
\ \ , \label{Schr}
\end{equation}

\noi under the Galilei transformations:
\bea
t'&=&t+b \nn\\
\vec{x}'&=&R\vec{x}+\vec{a}+\vec{V}t\label{galileo}\\
\vec{v}'&=&R\vec{v}+\vec{V}\;.\nn
\eea

\noi where $R$ represents rotations, $b\in \R$ and $\vec{a}\in {\R}^3$ time and space 
translations, respectively, and $\vec{V}\in {\R}^3$ galilean boosts. 

Equation (\ref{Schr}) acquires an extra term,

\begin{equation}
i\hbar\frac{\partial \,}{\partial t'}\Psi+i\hbar
\vec{V}\cdot\frac{\partial \Psi}{\partial \vec{x}'}
  =-\frac{\hbar^2}{2m}{\nabla'}^2\Psi \;,  \label{Schrp}
\end{equation}

\noi which can be compensated only by also transforming the wave function. Allowing 
for a non-trivial phase factor in front of the transformed wave function, of 
the form

\begin{equation}
\Psi'=e^{\frac{im}{\hbar}(\vec{V}\cdot R\vec{x}+\frac{1}{2}\vec{V}^2t)}\Psi\,, \label{Psi}
\end{equation}

\noi the Schr\"odinger equation becomes strictly invariant:

\begin{equation}
i\hbar\frac{\partial \,}{\partial t'}\Psi'=-\frac{\hbar^2}{2m}{\nabla'}^2\Psi' 
\end{equation}

The need for a transformation such as (\ref{Psi}) accompanying the space-time
transformation (\ref{galileo}) to accomplish full invariance strongly suggests
the adoption of a central extension of the Galilei group as the basic
(quantum-mechanical) space-time symmetry
for the free particle \cite{Bargmann}. The constant $\hbar$ is required to 
keep the exponent in (\ref{Psi}) dimensionless.  

The successive composition of two transformations in the extended Galilei 
group $\tilde{G}$ immediately leads to the group law:
\begin{eqnarray}
b''&=&b'+b \nn \\
\vec{a}''&=&\vec{a}'+R(\vec{\epsilon'})\vec{a}+\vec{V}'b \nn \\
\vec{V}''&=&\vec{V}'+R(\vec{\epsilon'})\vec{V} \\
\vec{\epsilon''}&=&\sqrt{1-\frac{\vec{\epsilon'}^2}{4}}\vec{\epsilon'}+
   \sqrt{1-\frac{\vec{\epsilon'}^2}{4}}\vec{\epsilon}+\frac{1}{2}\vec{\epsilon'}\wedge\vec{\epsilon}\nn\\
 e^{i\phi''}&=&e^{i\phi'}e^{i\phi}
e^{\frac{im}{\hbar}[\vec{V}'R'\vec{a}+\frac{1}{2}b\vec{V}'^2]}\;, \nn
\end{eqnarray}

\noi where $e^{i\phi}\in U(1)$, and we have made the
 rotation parameters $\vec{\epsilon}\in {\R}^3$ explicit, which are restricted to
$2\hbox{sin}\frac{\chi}{2}=|\vec{\epsilon}|,\,\chi$ being the rotation angle. 

\medskip

\subsection{Semi-invariance in Classical Mechanics}

\medskip

This phenomenon of extending the space-time symmetry, although
conceptually pure quantum mechanics, can also be recast within a (semi-)classical
formalism, by requiring the simultaneous extension of the classical
phase space by a new variable $\phi$, transforming in a non-trivial way
under the $U(1)$-extended symmetry group. The need for such an extension
is motivated by the lack of strict invariance of the Poincar\'e-Cartan
form associated with the free particle 
($H=p_i\stackrel{.}x^i-L=\frac{\vec{p}^2}{2m}$), 
\be
\Theta_{PC}\equiv p_idx^i-Hdt=(\frac{\!\partial L}{\partial\!\stackrel{.}{x}^i}(dx^i-
\!\stackrel{.}{x}^idt)+Ldt)= p_idx^i-\frac{\vec{p}^2}{2m}dt \label{PC}
\ee

\noi under the Galilei group. In fact, it is left only semi-invariant by the infinitesimal transformations
associated with (\ref{galileo})  in the sense that the Lie derivative
of $\Theta_{PC}$ with respect to those generators is the
differential of a function not necessarily zero:  
\be
\begin{array}{rclrclrclrcl}
X_b&=&\frac{\!\partial}{\partial t} & &\Rightarrow& & L_{X_b}\Theta_{PC}&=&0\\
X_{\vec{a}}&=&\frac{\!\partial}{\partial \vec{x}} & &\Rightarrow& & L_{X_{\vec{a}}}\Theta_{PC}&=&0\\
X_{\vec{V}}&=&t\frac{\!\partial}{\partial \vec{x}}+m\frac{\!\partial}{\partial \vec{p}}& &\Rightarrow& & L_{X_{\vec{V}}}\Theta_{PC}&=&d(m\vec{x})\\
X_{\vec{\epsilon}}&=&\vec{x}\wedge\frac{\!\partial}{\partial \vec{x}}+
   \vec{p}\wedge\frac{\!\partial}{\partial \vec{p}}& &\Rightarrow& & L_{X_{\vec{\epsilon}}}\Theta_{PC}&=&0
\end{array}\label{semi}
\ee

The pathology of semi-invariance is parallel to the absence of a clean
quotient by the equations of motion. Let us see in some detail the quotient
process in going to the solution manifold. In the Cartan formalism the
trajectories of a general physical system are the orbits of the kernel of
$d\Theta_{PC}$:
\bea 
\Theta_{PC}&\equiv& p_idx^i-Hdt \\
\Omega &\equiv&d\Theta_{PC}=dp_i\wedge dx^i-\frac{\partial H}{\partial x^i}
    dx^i\wedge dt-\frac{\partial H}{\partial p_i}dp_i\wedge dt\;.\nn
\eea

\noi $\Omega$ has a one-dimensional kernel generated by $X^H\in \hbox{Ker}\, d\Theta_{PC}$
such that $dt(X^H)=1$,

\be
X^H=\frac{\partial \;}{\partial t}+\frac{\partial H}{\partial p_i}\frac{\partial \;}
     {\partial x^i}-\frac{\partial H}{\partial x^i}\frac{\partial \;}{\partial p_i}\,,\label{XH}
\ee

\noi and the associated equations of motion are the Hamilton equations:

\bea
\frac{dt}{d\tau}&=&1\nn\\
\frac{dx^i}{d\tau}&=&\frac{\partial H}{\partial p_i}\\
\frac{dp_i}{d\tau}&=&-\frac{\partial H}{\partial x^i}\nn
\eea

The vector field $X^H$ defines a one-parameter group  which 
divides the {\it space of movements} $\R\times {\R}^3\times {\R}^3$,
parametrized by $(t,\;\vec{x},\;\vec{p})$, into classes, and
$M\equiv \{\R\times {\R}^3\times {\R}^3\}/X^H$ constitutes the
symplectic {\it phase space} of the system characterized 
by the Hamiltonian $H$; the symplectic form is obtained by the projection of $\Omega$. 
The change of variables under which the equations of motion on the quotient
become trivial is the Hamilton-Jacobi transformation. 
For the example $H=\frac{\vec{p}^2}{2m}$, corresponding to a free particle, this
transformation is:

\be
\left\{\ba{l}x^i=\frac{P^i}{m}\tau+K^i\\ p_i=P_i\\ t=\tau \ea \right.
\Longleftrightarrow \left\{\ba{l}K^i=x^i-\frac{p^i}{m}t\\ P_i=p_i\\ 
   \tau=t \ea \right.\;,\label{HJ}
\ee

\noi where the constants of motion $K^i,P_j$ parametrize the solution 
manifold $M$. However, the form $\Theta_{PC}$ goes to the quotient
except for a total differential:

\bea 
\Theta_{PC}&\rightarrow &P_idK^i+d(\frac{\vec{P}^2}{2m}\tau)\\
\omega &=&dP_i\wedge dK^i\nn
\eea 

\medskip

\subsection{Poisson algebra realization}

\medskip

Another equivalent Analytical-Mechanics breakdown claiming a
``generalization" is the unfair relationship between the Lie brackets of
basic symmetries and the corresponding Poisson brackets of the associated
Noether invariants. The symplectic form \cite{Godbillon,Abraham,Marsden} is a 
skew-symmetric ``metric" and defines an isomorphism
$\omega^\flat:{\cal X}(M)\leftrightarrow \Lambda^1(M)$ between the vector 
space of vector fields on $M$ and that of one-forms on $M$,

\be
X\in{\cal X}(M)\longmapsto\omega(X,\cdot)\equiv i_X\omega\in\Lambda^1(M)\,,
\ee

\noi associating a bracket $\{,\}$ on $\Lambda^1(M)$ with the Lie bracket of 
vector fields. In particular, given functions $f,g\in C^\infty(M)$, their 
differentials are associated with {\it Hamiltonian vector fields
$X_f,X_g$} \cite{footnote1}. This permits
the definition of a Poisson bracket between functions, rather than one-forms, 
but this time the correspondence $\{\;,\,\}\rightarrow [\;,\,]$,

\be
\{,\}:f,g\longmapsto\{f,g\}\,\,/\,\,d\{f,g\}=-i_{[X_f,X_g]}\omega\;,
\ee

\noi is no longer an isomorphism because constant functions have trivial
Hamiltonian vector fields. In particular, with regard to the example $H=\frac{\vec{p}^2}{2m}$,
and considering $K^i,P_j$ as the basic coordinates for $M$, we find:

\be
\{K^i,P_j\}=\delta^i_j\cdot 1\longmapsto [X_{K^i},X_{P_j}]=0
\ee

\noi that is, a Lie algebra homomorphism whose kernel is the central subalgebra of
constant functions, $\R$, generated by $1$. It is easy to realize that
$X_{K^i},X_{P_j}$ are nothing other than the generators $X_{V^i},
X_{a^j}$, respectively, of the action (\ref{galileo}) of the unextended
Galilei group on the space of movements, written on the solution manifold
\cite{footnote2}.

The extension of phase space is required to represent faithfully the
(classical) Poisson algebra by means of the generators of the extended
symmetry as first-order differential operators, a fact that constitutes the Bohr-Sommerfeld
approximation to Quantization or Prequantization, in the language of Geometric
Quantization \cite{Souriau,Kostant,Sniatycki,Woodhouse}. Let us 
look at the preliminary steps towards the geometric attempts at quantization. 

\medskip

\subsection{Geometric and group approach to quantization}

\medskip

The existence of a non-trivial kernel in the correspondence between
functions and Hamiltonian vector fields is an essential failure to
the naive geometric approach to quantization $\;\;\hat{}:f\mapsto \hat{f}\equiv
X_f$, which would associate the trivial operator to any constant. The 
simplest way of avoiding this problem consists of enlarging phase space 
(and/or movements space) with one extra variable providing one extra component to $X_f$, and 
generalizing accordingly the equation $i_{X_f}d\Theta_{PC}=-df$ so as to get a 
non-trivial new component even though $f$ is a constant. On a quantum 
manifold $P$, locally isomorphic to $M\times S^1$, with connection form 
$\Theta$ such that the curvature two-form ($d\Theta$) coincides with $d\Theta_{PC}$, the 
equation above can be replaced by the set of equations \cite{Souriau}:
\bea
i_{\tilde{X}_f}d\Theta &=& -df\nn\\
i_{\tilde{X}_f}\Theta &=& f \label{ixdteta}
\eea

\noi generalizing this way the quantization map which now reads (except perhaps
for a minus sign) 

\be
\hat{}:f\longmapsto i\tilde{X}_f
\ee 

\noi Note that Eq. (\ref{ixdteta}) inmediately implies the strict invariance
of $\Theta$ under $\tilde{X}_f$:

\be
L_{\tilde{X}_f}\Theta=di_{\tilde{X}_f}\Theta+i_{\tilde{X}_f}d\Theta=df-df=0
\ee

 Locally, we can write $\Theta=\Theta_{PC}+\frac{dz}{iz},\;z=e^{i\Phi}\in S^1$ and then
$\tilde{X}_f=X_f+[f-\Theta_{PC}(X_f)](iz\frac{\partial \;}{\partial z}-
iz^*\frac{\partial \;}{\partial z^*})$, and we 
immediately see that (\ref{ixdteta}) has a unique solution associating the
fundamental (vertical) vector field $\Xi\equiv iz\frac{\partial
\;}{\partial z}-iz^*\frac{\partial \;}{\partial
z^*}=\frac{\!\partial}{\partial \Phi}$, dual to $\frac{dz}{iz}$, with
the unity of $\R$.

The quantization map $\;\hat{}\;$ is now an isomorphism between the Poisson 
algebra on $M$ and the Lie subalgebra of vector fields on $P$ that are solutions
to (\ref{ixdteta}). For the basic functions, we have:
\be
\{K^i,P_j\}=\delta^i_j\cdot 1 \longleftrightarrow [\tilde{X}_{K^i},
      \tilde{X}_{P_j}]=\delta^i_j\cdot \Xi\;.
\ee

\noi It is again easy to realize that, in the case of the free particle,
the operators $\tilde{X}_{K^i}, \tilde{X}_{P_j}$ are nothing other than
the generators $\tilde{X}_{V^i}, \tilde{X}_{a^j}$ of the action
(\ref{galileo}) and (\ref{Psi}) on the extended space of movements (with
$\zeta\equiv e^{i\phi}\in S^1$) of the extended
Galilei group, that is to say, 
\bea
\tilde{X}_b&=&X_b\nn\\
\tilde{X}_{\vec{a}}&=&X_{\vec{a}}\\
\tilde{X}_{\vec{V}}&=&X_{\vec{V}}-\frac{1}{\hbar}m\vec{x}\frac{\!\partial}{\partial
   \phi}\nn\\
\tilde{X}_{\vec{\epsilon}}&=&X_{\vec{\epsilon}}\nn
\eea

\noi written on the solution manifold. 

To pass to the solution manifold, we must take into account the
evolution of the new variable $\zeta\in U(1)$ . In fact, the equations
of motion in the extended (by $U(1)$) movement
space are given by the vector field $\check{X}$ in the kernel of $d\Theta$ and
$\Theta$, simultaneously, satisfying $dt(X)=1$. Locally, and for the choice
$\Theta_{PC}=p_idx^i-Hdt$ we find:
\be
\check{X}^H=X^H+\{H-p_i\frac{\partial H}{\partial p_i}\}\Xi
\ee

\noi For the free particle, $\check{X}^H$ provides the following new
equation, to be added to de Hamilton-Jacobi set (\ref{HJ}),
\be
\zeta=ze^{-\frac{i}{\hbar}\frac{\vec{P}^2}{2m}t}\;\Longleftrightarrow\;z=\zeta
e^{\frac{i}{\hbar}\frac{\vec{p}^2}{2m}t}\label{HJseta}
\ee

Now the form $\Theta$,
originally written in the space of movements as
\be
\Theta=p_idx^i-\frac{\vec{p}^2}{2m}dt+\hbar\frac{d\zeta}{i\zeta}\label{Tetalibre}
\ee

\noi goes to the quotient, by applying the extended Hamilton-Jacobi
transformation (\ref{HJ}) and (\ref{HJseta}), giving
\be 
\Theta=P_idK^i+\hbar\frac{dz}{iz}\label{Teta} 
\ee

The space of wave functions $\Psi$ is constituted by the complex functions on $P$
that satisfy the $U(1)$-equivariance condition, turning $\Psi$ into a section
of the principal bundle $P\rightarrow M$ \cite{Nomizu}:

\be 
\Sigma\tilde{\Psi}=i\tilde{\Psi} \longleftrightarrow \tilde{\Psi}(K,P,z)=z\Psi(K,P)\,,
\ee

\noi on which the vector fields $\tilde{X}_f$ act, defining the {\it pre-quantum operators}.

Unfortunately, the quantization map $\;\hat{}\;$ is faithful but not irreducible
as a representation of the Lie algebra of classical functions. At this 
prequantization level, we are able to reproduce only the
Bohr-Sommerfeld-Wilson quantization rules \cite{Woodhouse}. We 
know that this representation is reducible because of the existence of
non-trivial operators commuting with the basic quantum generators
$\hat{K}^i\equiv i\hbar\tilde{X}_{K^i},\hat{P}_j\equiv
-i\hbar\tilde{X}_{P_j}$. In fact, thinking of the simplest case, that of
the free particle for example, and adopting for $\Theta$ the local
expression (\ref{Teta}), we get the following basic operators acting on
the untilded wave functions $\Psi$: 

\bea
\hat{K}^i &=&i\hbar\frac{\partial \; }{\partial P_i}+K^i \nn\\
\hat{P}_j &=&-i\hbar\frac{\partial \; }{\partial K^j}
\eea

\noi and it is clear that the operators $\check{K^i}\equiv\frac{\partial \; }{\partial P_i}$ 
do commute with them.

True quantization requires that all non-trivial operators commuting with basic quantum 
generators should be trivialized. We must then impose a maximal set of mutually compatible 
conditions in the form $X\Psi=0$, for $X$ in some maximal
vector space called {\it polarization}. For instance, in the example above the 
operator $\frac{\partial \;}{\partial P_i}$ would be trivial had we imposed
the polarization condition $\frac{\partial \;}{\partial P_i}\Psi=0\rightarrow
\Psi\neq\Psi(P)$. Finding a polarization, however, is  a non-trivial task in 
general, because two polarization conditions $\hat{a}\Psi=0,
\hat{b}\Psi=0$ are inconsistent if $[\hat{a},\hat{b}]=\hat{1}$ and once
a certain polarization has been imposed, the set of physical operators
that preserve the polarization is severely restricted. Even
more, the existence of an invariant polarization, i.e. a polarization
preserved by the basic operators, is by no means guaranteed.

A stylish and even practical (at least for fundamental systems) solution
to this and other problems comes from the structure itself of the (classical)
Poisson algebra seen as a fundamental symmetry of the physical system to
be quantized or, more precisely, the (quantum) physical system to be
studied. Looking again at the quantum symmetry of the free particle, the
extended Galilei group (\ref{galileo}), we can adopt its structure as
the basic block to provide the whole physical system, including the
space and time where the evolution takes place \cite{23,Ramirez}. 
The chief idea is to replace the quantum manifold $P$ with a Lie group 
$\tilde{G}$ bearing the structure of a principal bundle with structure
group $U(1)$ and a connection $1$-form $\Theta$ selected from the
canonical invariant forms on the group. The simplest, though 
sufficiently broad, example of such a class of Lie groups is the case of a 
central extension of a given group $G$ by $U(1)$ \cite{Bargmann}. In
this situation, the central extensions are parametrized by the second
co-homology group of $G$ in $U(1)$, $H^2(G,U(1))$, and the
coordinates in this (vector) space are associated with fundamental
constants such as the mass \cite{Souriau} or the electric charge (see
Sec. IV). Each co-homology constant is in turn associated with a Lie
subalgebra of even dimension (in fact, a symplectic vector space) which
will provide a set of canonically conjugate pairs of operators. We shall
call them {\it dynamical} quantities, or the corresponding parameters in the
group (classical) dynamical variables. 

The virtues of working on a Lie group are multiple, but, for the
time being, let us point out that of possessing two sets of natural,
mutually commuting operators; that is, the left- and right-invariant
vector fields, the latter of which can provide a unitary representation
to be reduced by polarization conditions imposed by a subalgebra of the
former. With this choice, the connection $1$-form $\Theta$ is the $U(1)$
component of the left-invariant canonical $1$-form on the group, which
is automatically invariant under the right-invariant vector fields. The
only apparent drawback of doing so is that the quotient $\tilde{G}/U(1)$
is not necessarily a symplectic manifold since the curvature $2$-form
$d\Theta$ may have a non-trivial kernel. However, this apparent problem
is solved by including in the polarization conditions, formulated in
terms of a maximal horizontal subalgebra ${\cal P}$ of left-invariant vector
fields, the subalgebra  of (left-invariant) vector fields ${\cal
G}_\Theta$ generating the characteristic module of $\Theta$, i.e.  
$\hbox{Ker}\;\Theta\cap\hbox{Ker}d\Theta$. 

It should be stressed that, far from being a drawback, working on
the pre-contact manifold $\tilde{G}$ instead of a proper quantum
manifold $P$ allows us to deal with quantum systems without classical
limit. In fact, the trajectories of the vector fields in the
characteristic subalgebra generalize the classical motion, and the
solution of the corresponding equations can be bypassed  by including
this subalgebra in the Polarization as generalized Schr\"odinger equations.

\medskip

\section{Cartan-like version of the Minimal-Coupling Principle: the
electromagnetic and non-relativistic gravitational forces}
 
\medskip

Once the symmetry of the free particle has been posed through the strict
invariance of the corresponding extended Poincar\'e-Cartan (or quantization) form
$\Theta$ (\ref{Tetalibre}) under the action of extended Galilei group, we
may postulate the requirement of invariance of a generalized $\Theta$
under the Galilei group (non-centrally) extended by the ``local" group
$U(1)(\vec{x},t)$ (of local phase transformations
$e^{i\phi(\vec{x},t)}$). 
This requirement, along with the minimal substitution
in $\Theta$ to achieve strict invariance, constitutes the Cartan-like
version of the Minimal Coupling Principle for the $U(1)$ rigid symmetry
and will lead to the motion of a particle in the presence of an
electromagnetic field.

Let us consider the Lie algebra $\tilde{\cal G}$ of the centrally 
extended Galilei group $\tilde{G}$ (only non-zero commutators): 
%
\begin{equation}
\begin{array}{rclrclrcl}
{}[\tilde{X}_{V^i},\,\tilde{X}_b]&=&\tilde{X}_{a^i} &
[\tilde{X}_{V^i},\,\tilde{X}_{a^j}]&=&\frac{m}{\hbar}\delta_{ij}\tilde{X}_\phi & & & \\ 
&&&&&&&&\\
{}[\tilde{X}_{\epsilon^i},\,\tilde{X}_{\epsilon^j}]&=&{\epsilon_{ij.}}^{\!k}\tilde{X}_{\epsilon^k} \;\;\; & 
[\tilde{X}_{\epsilon^i},\,\tilde{X}_{V^j}]&=&{\epsilon_{ij.}}^{\!k}\tilde{X}_{V^k} \;\;\;&
[\tilde{X}_{\epsilon^i},\,\tilde{X}_{a^j}]&=&{\epsilon_{ij.}}^{\!k}\tilde{X}_{a^k}
\end{array}\label{Galileo}
\end{equation}
 
\noi which, as mentioned above, leaves strictly invariant the extended
Poincar\'e-Cartan form $\Theta=p_idx^i-\frac{\vec{p}^2}{2m}+\hbar d\phi$, that is,
$L_{\tilde{X}_a}\Theta=0, \forall \tilde{X}_a\in\tilde{\cal G}$.  
 
Local $U(1)$ transformations generated by $f\otimes X_\phi$, $f$ being a real function 
$f(\vec{x},t)$, are incorporated into the scheme by adding to (\ref{Galileo}) 
the extra commutators \cite{footnote3}:
%
\be 
[\tilde{X}_a,\,f\otimes X_\phi]=(L_{\tilde{X}_a}f)\otimes X_\phi\label{fconmutador} 
\ee 

\noi The Lie derivative of (\ref{Tetalibre}) with respect to $f\otimes
X_\phi$ gives:
\[ L_{f\otimes X_\phi}\Theta=d(i_{fX_\phi}\Theta)+i_{fX_\phi}d\Theta=df \]

\noi Keeping the strict invariance requires modifying $\Theta$ 
by adding a connection term $\Gamma=\Gamma_idx^i+\Gamma_0dt$ whose
components transform under 
$U(1)(\vec{x},t)$ as the space-time gradient of the function $f$
\cite{International}. Additional conditions on $\Gamma$ will be obtained by
requiring strict invariance of $\Theta'\equiv\Theta+\Gamma$ under the complete group. The
generators of the action of the whole group on the variables
($t,\,\vec{x},\,\vec{p},\,\vec{\Gamma},\,\Gamma_0,\phi$) are:
\bea
\tilde{X}_b&=&\frac{\!\partial}{\partial t}\nn\\
\tilde{X}_{\vec{a}}&=&\frac{\!\partial}{\partial \vec{x}}\\
\tilde{X}_{\vec{V}}&=&t\frac{\!\partial}{\partial \vec{x}}+m\frac{\!\partial}
    {\partial \vec{p}}+\vec{\Gamma}\frac{\!\partial}{\partial \Gamma_0}+
    \frac{m}{\hbar}\vec{x}\frac{\!\partial}{\partial\phi}\nn\\
\tilde{X}_{\vec{\epsilon}}&=&\vec{x}\wedge\frac{\!\partial}{\partial \vec{x}}+
   \vec{p}\wedge\frac{\!\partial}{\partial\vec{p}}+
   \vec{\Gamma}\wedge\frac{\!\partial}{\partial\vec{\Gamma}}\nn\\
f\otimes X_\phi&=&-\vec{\nabla}
   f\frac{\!\partial}{\partial\vec{\Gamma}}+\frac{\partial f}{\partial t}
   \frac{\!\partial}{\partial\Gamma_0}-\frac{f}{\hbar}\frac{\!\partial}{\partial\phi}\nn
\eea

\noi Then, the infinitesimal conditions
\[ L_X(\Theta')=0\]

\noi implies the following finite transformation properties of the
   components of $\Gamma$:
\bea
\vec{\Gamma}'&=&R\vec{\Gamma}\nn\\
\Gamma_0'&=&\Gamma_0+\vec{V}\cdot R\vec{\Gamma}\label{Gammas}
\eea

\noi under a rotation and a boost, and
\bea
\vec{\Gamma}'&=&\vec{\Gamma}+\vec{\nabla}f\label{gradiente}\\
\Gamma_0'&=&\Gamma_0+\frac{\partial f}{\partial t}\nn
\eea

\noi under an element of $U(1)(\vec{x},t)$.

Let us now compute the simultaneous kernel of $\Theta'$ and
$d\Theta'$ and write the equations of motion, rewriting the connection $\Gamma$ as $\Gamma\equiv
qA_idx^i-qA_0dt$. We then have (omitting the accent over $\Theta'$)
\be 
\Theta=m\vec{v}\cdot d\vec{x}-\frac{1}{2}m\vec{v}^2dt+q\vec{A}\cdot
   d\vec{x}-qA^0dt+\hbar d\phi \label{tetaelectrificante} 
\ee 
\be
\check{X}=\frac{\!\!\partial}{\partial t}+\vec{v}\cdot\frac{\!\!\partial}{\partial\vec{x}}+ 
  \frac{q}{m}\left[\left(\frac{\partial A_j}{\partial x^i}-\frac{\partial A_i}{\partial x^j} 
  \right)v^j-\frac{\partial A^0}{\partial x^i}-\frac{\partial A_i}{\partial t}\right] 
  \frac{\!\!\partial}{\partial v_i}-\frac{1}{\hbar}[\frac{1}{2}m\vec{v}^2+
  q(\vec{v}\cdot\vec{A}-A^0)] 
  \frac{\!\!\partial}{\partial\phi},\label{Xmovimiento} 
\ee 

\noi It states the equations of motion of a charged particle of charge
$q$ in an electromagnetic field:
\bea 
\frac{d\vec{x}}{dt}&=&\vec{v}\label{ecuElec}\\ 
m\frac{d\vec{v}}{dt}&=&q\left[\vec{v}\wedge(\vec{\nabla}\wedge\vec{A})-\vec{\nabla}A^0- 
     \frac{\partial\vec{A}}{\partial t}\right]\nn\\
\frac{d\phi}{dt}&=&-\frac{1}{\hbar}\left(\frac{\vec{p}{}^2}{2m}-
      \frac{q}{m}\vec{A}\cdot\vec{p}\right)\nn   
\eea 

\medskip

The Minimal-Coupling Principle can also be applied to the case of
Newtonian Gravity by requiring the space and time translation parameters to
depend on time. Unlike the electromagnetic gauge principle, which can be
directly extended to the relativistic situation, relativistic gravity is
much more involved and will be analysed in Sec. V, mixed with
electromagnetism as our central task.

For brevity, let us consider the $1+1$-dimensional
case. Starting with the unextended Galilei group, we promote 
the space and time translations to local in the more economical way; that is, turning
the corresponding group parameters into functions of time. The gauge
algebra is then
\bea
[f(t)\otimes X_b,\,X_b]&=&-\frac{\partial f}{\partial t}\otimes X_b\nn \\
\left[f(t)\otimes X_b,\,X_a\right]&=&0\label{Galileogauge} \\
\left[f(t)\otimes X_b,\,X_V\right]&=&f(t)\otimes X_a \nn
\eea

According to the minimal prescription, we introduce a conection
$\Gamma\equiv hdt$ to be added to the free Poincar\'e-Cartan form,
and an extra component in $\frac{\!\partial}{\partial h}$ to the
Galilei generators for non-trivial realization of the current algebra
(\ref{Galileogauge}). Then the semi-invariance of
\be
\Theta_{PC}'=pdx-\frac{p^2}{2m}dt+hdt\label{Tetagrav}
\ee
%

\noi under the current algebra
(\ref{Galileogauge}) fixes the new components in $h$, so that the
complete expression of the Lie algebra of the {\it unextended} Galilei group
with local (depending only on time) space and time translation subgroup becomes:
\bea
f(t)\otimes X_b&=&f\frac{\!\partial}{\partial t}+(\frac{p^2}{2m}-h)\frac{df}{dt}
   \frac{\!\partial}{\partial h}\nn\\
f(t)\otimes X_a&=&f\frac{\!\partial}{\partial x}-p\frac{df}{dt}\frac{\!\partial}
   {\partial h}\nn\\
X_V&=&t\frac{\!\partial}{\partial x}+m\frac{\!\partial}{\partial p}\label{stcurrent}\\
f(t)\otimes X_{h}&=&f\frac{\!\partial}{\partial h}\nn
\eea

\noi where we have had to introduce the new local generator $f(t)\otimes
X_{h}$ in order to close the current algebra. The new generator
also leaves the form (\ref{Tetagrav}) semi-invariant. Strict invariance
is now achieved by adding to generators $X$ new components in
$\frac{\!\partial}{\partial\phi}$ with coefficients
equal to $-g$, $g$ being such that $i_Xd\Theta'=dg$. This results in:
\bea
\widetilde{f(t)\otimes X_b}&=&f\frac{\!\partial}{\partial t}+(\frac{p^2}{2m}-h)\frac{df}{dt}
   \frac{\!\partial}{\partial h}-\frac{f}{\hbar}(\frac{p^2}{2m}-h)
  \frac{\!\partial}{\partial\phi}\nn\\
\widetilde{f(t)\otimes X_a}&=&f\frac{\!\partial}{\partial x}-p\frac{df}{dt}\frac{\!\partial}
   {\partial h}+\frac{1}{\hbar}fp\frac{\!\partial}{\partial\phi}\nn\\
\tilde{X}_V&=&t\frac{\!\partial}{\partial x}+m\frac{\!\partial}{\partial p}-
   \frac{m}{\hbar}(x-\frac{p}{m}t)\frac{\!\partial}{\partial\phi}\label{extstcurrent}\\
\widetilde{f(t)\otimes X_{h}}&=&f\frac{\!\partial}{\partial h}-
    \frac{f}{\hbar}\frac{\!\partial}{\partial\phi}\nn
\eea 

\noi The reason for $U(1)$-extending after gauging is clear: both
processes definitely do not commute. Proceeding the other way round
leads, precisely, to the new results of Sec. V.

The Cartan-like equations associated with (\ref{Tetagrav}) lead directly
to the Newtonian gravity equations if we identify $h$ with the
gravitational potential. This potential can be related to the
component $g_{00}$ of a metric in the Newtonian limit of General Relativity 
via the expression $g_{00}\approx 1+h$.

\medskip

\section{Group Approach to the Quantization of a particle moving in an
electromagnetic field}

\medskip

In this and the next section, we shall adopt the GAQ formalism as a
generalization of the geometrical approach to 
Quantum Mechanics, although we shall be  interested, for now, primarily
in the classical equations of motion. As mentioned above, we
seek to reproduce any dynamical or kinematical quantity or variable out
of a Lie group so that notation such as $t,\vec{x},\vec{p}$, etc. will
refer to group variables (although directly identifiable with
``physical" ones once the equations of motion are written). 

Let us start by exponentiating the algebra
(\ref{Galileo})+(\ref{fconmutador}), originally performed on a given
movement space, in order to arrive at an abstract Lie group from
which to obtain all physical structures. This algebra is infinite-dimensional
but, for real analytic functions 
$f$, the dynamical-variable content of it, in the sense of
Sec. {\bf 2.4}, is addressed by the (co-homological) structure 
of the finite-dimensional subalgebra generated by $\tilde{\cal G}$ along
with those generators $f\otimes 
X_\phi$ with only linear functions, $t\otimes X_\phi$ and $x^i\otimes
X_\phi$, to be called $X_{A^0}$ and $X_{A^i}$, respectively. The rest of functions
contribute only to the characteristic (non-dynamical) subalgebra and can be
decoupled from the theory.  Let us call
$\tilde{G}_E$ this finite-dimensional group. It proves to be enough to describe the
dynamics of a particle moving in an electromagnetic field if we resort
to the trick (see below) of assuming an explicit dependence
$A^{\mu}=A^{\mu}(\vec{x},t)$ once the 1-form $\Theta$ will be found.
(This procedure is suggested by the possibility of writing an analytic
function in the form $f(\vec{x},t)=\phi+A_\mu(\vec{x},t)x^\mu$, where  $\phi=f(\vec{0},0)$.)
We shall not be involved here with the corresponding quantum field theory.

The group $\tilde{G}_E$ can be given the following group law which
extends that of the Galilei group (with parameters
$t,\vec{x},\vec{v}\equiv\frac{\vec{p}}{m}$ instead of
$b,\vec{a},\vec{V}$, respectively) and agree with the finite
transformation (\ref{Gammas}) and (\ref{gradiente}):
\bea
t''&=&t'+t\nn\\
\vec{x}''&=&\vec{x}'+R'\vec{x}+\vec{v}'t\nn\\
\vec{v}''&=&\vec{v}'+R'\vec{v}\nn\\
R''&=&R'R\label{GE}\\
\vec{A}''&=&\vec{A}'+R'\vec{A}\nn\\
A_0''&=&A_0'+A_0+\vec{v}'\cdot R'\vec{A}\nn\\
\zeta''&=&\zeta'\zeta e^{i\xi_m(g',g)}e^{i\xi_q(g',g)}\nn\\
\xi_m(g',g)&\equiv&-\frac{m}{\hbar}[\vec{v'}\cdot R'\vec{x}+\frac{1}{2}t{\vec{v'}}^2]\nn\\
\xi_q(g',g)&\equiv&-\frac{q}{\hbar}[\vec{A'}\cdot R'\vec{x}+t(\vec{v'}\cdot\vec{A'}-A_0')]\nn
\eea

\noi where $\xi_m(g',g)$ is a standard Bargmann-like cocycle  associated
with the Galilei (sub)group, in particular with the symplectic
submanifold of coordinates $(x^i,v_j)$, and $\xi_q(g',g)$ is a new cocycle,
parametrized by the electric charge, as we shall see, associated
with the symplectic submanifold of coordinates $(x^i, A_j)$. Both
satisfy the cocycle conditions:
\bea
\xi(g',g)+\xi(g'*g,g'')&=&\xi(g',g*g'')+\xi(g,g'')\nn\\
\xi(0,g)=\xi(g',0)&=&0\nn
\eea

\noi intended to mantain the structure of group law after the
extension. Since both cocycles are associated with intersecting symplectic
submanifold ($\vec{x}$ is in both) we should expect a mixed
momentum variable conjugated to $\vec{x}$, to be identified with the
minimally coupled momentum. This is the essence of  Minimal Coupling
in GAQ.  

>From (\ref{GE}) we derive left- and right-invariant vector fields and
from the former the $\zeta$-component of the left-invariant canonical
$1$-form:
\bea
{\tilde{X}_t}^L&=&\frac{\!\partial}{\partial t}+\vec{v}\cdot
   \frac{\!\partial}{\partial\vec{x}}-\frac{1}{\hbar}[\frac{1}{2}m\vec{v}^2+
   q(\vec{v}\cdot\vec{A}-A_0)]\Xi\nn\\
{\tilde{X}_{\vec{x}}}^L&=&R\left(\frac{\!\partial}{\partial\vec{x}}-\frac{1}{\hbar}[m\vec{v}+q\vec{A}]\Xi\right)\nn\\
{\tilde{X}_{\vec{v}}}^L&=&R\left(\frac{\!\partial}{\partial\vec{v}}\right)\label{LGE}\\
{\tilde{X}_{\vec{\epsilon}}}^L&=&\sqrt{1-\frac{\vec{\epsilon}^2}{4}}\frac{\!\partial}
    {\partial\vec{\epsilon}}-\frac{1}{2}\vec{\epsilon}\wedge\frac{\!\partial}
    {\partial\vec{\epsilon}}\nn\\
{\tilde{X}_{\vec{A}}}^L&=&R\left(\frac{\!\partial}{\partial\vec{A}}+\vec{v}\frac{\!\partial}
    {\partial\vec{A_0}}\right)\nn\\  
{\tilde{X}_{A_0}}^L&=&\frac{\!\partial}{\partial A_0}\nn\\
{\tilde{X}_\zeta}^L&=&i(\zeta\frac{\!\partial}{\partial\zeta}-\zeta^*\frac{\!\partial}
   {\partial\zeta^*})\equiv\frac{\!\partial}{\partial\phi}\equiv\Xi\nn
\eea

\medskip

\bea
{\tilde{X}_t}^R&=&\frac{\!\partial}{\partial t}\nn\\
{\tilde{X}_{\vec{x}}}^R&=&\frac{\!\partial}{\partial\vec{x}}\nn\\
{\tilde{X}_{\vec{v}}}^R&=&\frac{\!\partial}{\partial\vec{v}}+t\frac{\!\partial}
   {\partial\vec{x}}  \label{RGE}\\
{\tilde{X}_{\vec{\epsilon}}}^R&=&\sqrt{1-\frac{\vec{\epsilon}^2}{4}}\frac{\!\partial}
    {\partial\vec{\epsilon}}+\frac{1}{2}\vec{\epsilon}\wedge\frac{\!\partial}
    {\partial\vec{\epsilon}}\nn\\
{\tilde{X}_{\vec{A}}}^R&=&\frac{\!\partial}{\partial\vec{A}}-q\vec{x}\Xi\nn\\  
{\tilde{X}_{A_0}}^R&=&\frac{\!\partial}{\partial A_0}+qt\Xi\nn\\
{\tilde{X}_\zeta}^R&=&i(\zeta\frac{\!\partial}{\partial\zeta}-\zeta^*\frac{\!\partial}
   {\partial\zeta^*})\equiv\frac{\!\partial}{\partial\phi}\equiv\Xi\nn
\eea

\medskip

\be
\Theta\equiv\hbar\theta^{(\zeta)L}=m\vec{v}\cdot d\vec{x}-\frac{1}{2}m\vec{v}^2dt+
  q\vec{A}\cdot d\vec{x}-qA_0dt+\hbar\frac{d\zeta}{i\zeta}
\ee

The commutation relations of (let us say) left generators (omitting
rotations, which operate in the standard way) are:
%
\begin{equation}
\begin{array}{rclrclrcl}
{}[{\tilde{X}_t}^L,\,{\tilde{X}_{x^i}}^L]&=&0&[{\tilde{X}_t}^L,\,{\tilde{X}_{v^i}}^L]&=&
 -{\tilde{X}_{x^i}}^L & [{\tilde{X}_{x^i}^L}\,,{\tilde{X}_{v^j}}^L]&=&
 \frac{m}{\hbar}\delta_{ij}\Xi\\
&&&&&&&&\\
{}[{\tilde{X}_t}^L,\,{\tilde{X}_{A^i}}^L]&=&0
 &[{\tilde{X}_t}^L,\,{\tilde{X}_{A^0}}^L]&=&-\frac{q}{\hbar}\Xi &
 [{\tilde{X}_{x^i}^L},\,{\tilde{X}_{A^j}}^L]&
  =& \frac{q}{\hbar}\delta_{ij}\Xi \\
&&&&&&&&\\ 
{}[{\tilde{X}_{x^i}}^L,\,{\tilde{X}_{A^0}}^L]&=&0 \;\;\;&
 [{\tilde{X}_{v^i}}^L,\,
 {\tilde{X}_{A^i}}^L]&=&\delta_{ij}{\tilde{X}_{A^0}}^L
 \;\;\;&[{\tilde{X}_{v^i}}^L,\,
  {\tilde{X}_{A^0}}^L]&=&0\nn 
\end{array}\label{Electro}
\end{equation}

If we compute the characteristic module of $\Theta$,
i.e. $\hbox{Ker}d\Theta\cap \hbox{Ker}\Theta$ for $q=0$, as
corresponding to the free particle, we find that it is generated by a
left subalgebra
\be
{\cal
G}_\Theta|_{q=0}=<{\tilde{X}_t}^L,\;{\tilde{X}_{\vec{\epsilon}}}^L,\;
   {\tilde{X}_{\vec{A}}}^L,\;{\tilde{X}_{A_0}}^L>\;,
\ee

\noi leading to the trajectories (\ref{HJ}) and (\ref{HJseta}) for the
dynamical (symplectic) variables $\vec{x}$ and $\vec{p}$, and additional ones for the
kinematical (non-symplectic) variables $\vec{\epsilon}, \vec{A}, A_0$ which decouple from
the theory. The quotient of $\tilde{G}_E/U(1)$ by the generalized
equations of motion is a symplectic manifold (the solution manifold) of
dimension $3+3$. However, for non-zero $q$, we have
\be
{\cal G}_\Theta=<{\tilde{X}_{\vec{\epsilon}}}^L,\;{\tilde{X}_{\vec{A}}}^L-\frac{q}{m}
   {\tilde{X}_{\vec{v}}}^L>\;,
\ee

\noi leading to a symplectic manifold of dimension $4+4$. We should note
that no time evolution appears as an equation of motion. This is because
the electromagnetic cocycle lends dynamical character to
$t$ as conjugate to $A_0$. The only characteristic vector field, apart
from ${\tilde{X}_{\vec{\epsilon}}}^L$, which again simply decouples the
variables $\vec{\epsilon}$, is the one defining the Minimal Coupling. In
fact, the Noether invariants $i_{\tilde{X}}\Theta$ are:
\bea
i_{{\tilde{X}_t}^R}\Theta&=&-(\frac{1}{2}m\vec{v}^2+qA_0)\nn\\
i_{{\tilde{X}_{\vec{x}}}^R}\Theta&=&m\vec{v}+q\vec{A}\equiv\vec{P}\nn\\
i_{{\tilde{X}_{\vec{v}}}^R}\Theta&=&-m(\vec{x}-\vec{v}t)+q\vec{A}t\nn\\
i_{{\tilde{X}_{\vec{A}}}^R}\Theta&=&-q\vec{x}\nn\\
i_{{\tilde{X}_{\vec{x}}}^R}\Theta&=&qt\nn
\eea

\noi reproducing, in particular the ``canonical momentun"
$\vec{P}\equiv m\vec{v}+q\vec{A}$.

Real dynamics appear when we impose the ``constraint"
$A_{0,i}=A_{0,i}(\vec{x},t)$ on $\Theta$, whose characteristic module
turns out
now to be generated by $\check{X}$ in (\ref{Xmovimiento}), thus
reproducing the standard equations of motion. The trick of introducing
this constraint after the form $\Theta$ associated with $\tilde{G}_E$
has been computed  can in fact be justified in mathematical terms although at
the price of introducing an explicit infinite parametrization of the
field $A_\mu$ by means of, for instance, Fourier coefficients
$a_\mu(\vec{k}),\,a^*_\nu(\vec{k})$, and very importantly,
an extra space-time translation group associated with this field, let us
say $\chi_\mu$. That is, the general space-time position on which the field
lies, conceptually differs from the space-time position of the particle. 
Under these conditions a cocycle of this infinite-dimensional group 
can be introduced, contributing the form $\Theta$ with the term
$q\vec{A}(\vec{x},t)\cdot d\vec{x}-qA_0(\vec{x},t)dt$ where
\be
A_\mu(\vec{x},t)=\int{\frac{d^3\vec{k}}{2k^o}\{a_\mu(\vec{k})e^{-ik\cdot x}+
  a^*_\mu(\vec{k})e^{ik\cdot x}\}}\,.
\ee

\noi That is, the electromagnetic field evaluated on the trajectories of the
particle. Now, the abovementioned constraint proves to be as natural as
stating that, on a trajectory,  the particle sees the field $A_\mu$
evaluated on $x_\mu$ rather than on $\chi_\mu$. This precise construction, along with the (also
infinite-dimensional) cocycle providing dynamical content to the field
variables $a_\mu(\vec{k}),\,a^*_\nu(\vec{k})$ themselves, deserves a
separate work, which is in course. 

We should say to conclude this section that this study can be repeated
with the centrally extended Poincar\'e group $\tilde{P}$
\cite{Saletan,pseudo} (see also
\cite{Conforme}) by
promoting to {\it local} the $U(1)$ transformations and considering the 
finite-dimensional subgroup $\tilde{P}_E$ analogous to $\tilde{G}_E$.

\medskip

\section{Mixing the electromagnetic and gravitational forces}

\medskip

Let us now consider the gravitational interaction from our
group-theoretical viewpoint (we shall omit $\hbar$ in this
section). To this end, we start 
directly with the centrally extended Poincar\'e group $\tilde{P}$
and see how 
the fact that the translation generators produce the central term under
commutation with some other generators (boosts) plays a singular role in the 
relationship between local space-time translations and local $U(1)$ 
transformations. Symbolically denoting the generators of translations by
$P,\,P_0$, those of boosts by $K$ and the central one by $\Xi$, we find: 
\be 
[K,\,f\otimes P]\simeq(L_Kf)\otimes P+f\otimes (P_0+\Xi)\,.\label
{Poincarepseudo} 
\ee 
 
\noi This means that turning the translations into {\it local} symmetry also entails the
{\it local} nature of the $U(1)$ phase. 
We thus expect a non-trivial mixing of gravity and electromagnetism
into an infinite-dimensional electro-gravitational group. 
 
We shall follow steps identical to those given in the former example. In
turning the parameters of space-time translations {\it local}, we replace
$x^\mu$ with $x^\mu+h_{\mu\nu}(\vec{x},x^0)x^\nu$, write a
finite-dimensional algebra ${\cal P}_{EG}$ keeping only the linear part
of local space-time translations ($x^\mu+h_{\mu\nu}x^\nu$, with
constant $h_{\mu\nu}$), the generators of which will be called
$X_{h^{\mu\nu}}$, apply the GAQ formalism and impose the ``constraint" $h^{\mu\nu} 
=h^{\mu\nu}(\vec{x},x^0)$, on the symplectic submanifolds (solution
manifolds). However, the co-homological structure 
of this finite-dimensional electro-gravitational subgroup, $\tilde{P}_{EG}$ is
richer than that of $\tilde{P}_E$ and the exponentiation 
of the Lie algebra $\tilde{\cal P}_{EG}$ is by far more involved. 
As mentioned in the Introduction, we attempt only
a basic description of the new phenomenology which results from
the present revisited gauge principle, although we seek, apart from the
exact Lorentz force, an approximate expression
for the geodesic force in terms of the metric $g^{\mu\nu}\equiv\eta^{\mu\nu}+h^{\mu\nu}$. 
Thus, we shall resort only to what seems to be the basic co-homological (fundamental)
constants corresponding to the inertial mass $m$, the electric charge
$q$, the gravitational
mass $g$ and the  {\it mixing vertex} coupling constant $\kappa$. 

Let us write the algebra $\tilde{\cal P}_{EG}$ in an almost covariant way (the  
central extensions and induced deformations are necessarily non-covariant). To this end, we 
parametrize the Lorentz transformations with $\epsilon^{\mu\nu}$ as usual. The proposed explicit  
algebra is: 
%
\bea 
\left[{\tilde{X}_{x^\mu}}^L, {\tilde{X}_{\epsilon^{\nu\rho}}}^L\right]&=&-\eta_{\nu\mu}{\tilde{X}_{x^{\rho}}}^L+ 
\eta_{\rho\mu}{\tilde{X}_{x^{\nu}}^L}-(m+\kappa q)c(\eta_{\rho\mu}\delta^0_\nu- 
\eta_{\nu\mu}\delta^0_\rho)\Xi \nn \\ 
\left[{\tilde{X}_{x^\mu}}^L, {\tilde{X}_{h^{\nu\rho}}}^L\right]&=&-\eta_{\nu\mu}{\tilde{X}_{x^{\rho}}}^L- 
\eta_{\rho\mu}{\tilde{X}_{x^{\nu}}}^L+\left[2(g-mc)\eta_{0\mu}\delta^0_\nu\delta^0_\rho +mc(\eta_{\rho\mu}\delta^0_\nu+ 
\eta_{\nu\mu}\delta^0_\rho)\right]\Xi \nn \\ 
\left[{\tilde{X}_{x^\mu}}^L, {\tilde{X}_{A^{\nu}}}^L\right]&=&-q \eta_{\nu\mu}\Xi \nn \\ 
\left[{\tilde{X}_{\epsilon^{\mu\nu}}}^L,{\tilde{X}_{\epsilon^{\alpha\beta}}}^L\right]&=&-\eta_{\alpha\nu}{\tilde{X}_{\epsilon^{\mu\beta}}}^L+ 
\eta_{\beta\nu}{\tilde{X}_{\epsilon^{\mu\alpha}}}^L+\eta_{\alpha\mu}{\tilde{X}_{\epsilon^{\nu\beta}}}^L- 
\eta_{\mu\beta}{\tilde{X}_{\epsilon^{\nu\alpha}}}^L \nn \\ 
\left[{\tilde{X}_{\epsilon^{\mu\nu}}}^L,{\tilde{X}_{h^{\alpha\beta}}}^l\right]&=&-\eta_{\alpha\nu}{\tilde{X}_{h^{\mu\beta}}}^L- 
\eta_{\beta\nu}{\tilde{X}_{h^{\mu\alpha}}}^L+\eta_{\alpha\mu}{\tilde{X}_{h^{\nu\beta}}}^L+ 
\eta_{\mu\beta}{\tilde{X}_{h^{\nu\alpha}}}^L+  \\ 
& &\frac{1}{q}\{\kappa q
c\left(\eta_{\alpha\nu}\delta^{\rho}_{\beta} 
\delta^0_\mu- \eta_{\mu\alpha}\delta^{\rho}_{\beta}\delta^0_\nu+ 
\eta_{\nu\beta}\delta^{\rho}_{\alpha}\delta^0_\mu- \eta_{\mu\beta} 
\delta^{\rho}_{\alpha}\delta^0_\nu\right)- \nn\\
& & 2(g-mc)\left[(\eta_{\alpha\nu}\delta^{\rho}_{\beta} 
\delta^0_\mu- \eta_{\mu\alpha}\delta^{\rho}_{\beta}\delta^0_\nu+ 
\eta_{\nu\beta}\delta^{\rho}_{\alpha}\delta^0_\mu- \eta_{\mu\beta} 
\delta^{\rho}_{\alpha}\delta^0_\nu)\delta^\rho_0+\right.\nn\\
& &\left.\delta^0_\alpha\delta^0_\beta(\eta_{0\nu}\delta^0_\mu-\eta_{0\mu}\delta^\rho_\nu)\right]\}{\tilde{X}_{A^\rho}}^L\nn\\   
\left[{\tilde{X}_{\epsilon^{\mu\nu}}}^L,{\tilde{X}_{A^\rho}}^L\right]&=&-\eta_{\rho\nu}{\tilde{X}_{A^\mu}}^L+\eta_{\rho\mu}{\tilde{X}_{A^\nu}}^L  \nn \\ 
\left[{\tilde{X}_{h^{\mu\nu}}}^L,{\tilde{X}_{h^{\alpha\beta}}}^L\right]&=&-\eta_{\alpha\nu}{\tilde{X}_{\epsilon^{\mu\beta}}}^L- 
\eta_{\beta\nu}{\tilde{X}_{\epsilon^{\mu\alpha}}}^L-\eta_{\alpha\mu}{\tilde{X}_{\epsilon^{\nu\beta}}}^L- 
\eta_{\mu\beta}{\tilde{X}_{\epsilon^{\nu\alpha}}}^L+\nn \\ 
& &\frac{1}{q}\{-\kappa q c\left[\eta_{\alpha\nu} \delta^{0\rho}_{\beta\mu} 
+ \eta_{\beta\nu}\delta^{0\rho}_{\alpha\mu}
+ \eta_{\alpha\mu}\delta^{0\rho}_{\beta\nu}+\right.\nn\\
& &\left. \eta_{\beta\mu}\delta^{0\rho}_{\alpha\nu}
\right]+
2(g-mc)\left[\delta^0_\alpha\delta^0_\beta(\eta_{0\nu}\delta^\rho_\mu+\eta_{0\beta}\delta^\rho_\nu)-\delta^0_\mu\delta^0_\nu(\eta_{0\beta}\delta^\rho_\beta+\eta_{0\alpha}\delta^\rho_\beta)\right]\}{\tilde{X}_{A^\rho}}^L \nn \\ 
\left[{\tilde{X}_{h^{\mu\nu}}}^L,{\tilde{X}_{A^\rho}}^L\right]&=&-\eta_{\rho\nu}{\tilde{X}_{A^\mu}}^L-\eta_{\rho\mu}{\tilde{X}_{A^\nu}}^L \nn  
\eea 
 
\noindent where $\delta^{0\rho}_{\beta\mu}\equiv
\delta^0_\beta \delta^{\rho}_{\mu}-\delta^0_\mu\delta^{\rho}_{\beta}$ is the Kronecker tensor.

It bears mentioning that one of the central extension parameters,
actually $g$, is really free at the Lie algebra level but must
acquire the value $g=mc$ if the present theory is intended to
reproduce the standard disconnected electromagnetic and gravitational
forces for $\kappa=0$, i.e. when the constant responsible for the mixing
of both interactions is switched off. The appearance
of relationships between two co-homology constants, such as $g=mc$,
indicates the compatibility of further extensions of the algebra
${\cal P}_{EG}$ with the constants already introduced. Such further extensions could
generalize the present results.

This algebra must be exponentiated in order to have a group law from
which to compute left- and right-invariant vector field and the
quantization $1$-form $\Theta$, just as in the pure electromagnetic
example. As stated above, such a process is much more involved and, in
principle, a ``perturbative" algorithm is in order. We shall
resort to an approximation formula \cite{Formal} up
to a given order (order 3 in the fully relativistic case and 4 in the
non-relativistic limit given in \cite{electrograv}, although in the
latter the expressions obtained prove to be exact already at this order),
inspired in the theory of formal groups \cite{Serre}, which generalizes that
of Campbell-Hausdorff in the sense that it allows for expressions
which  more directly fit actual physical formulae (although the latter is
as well valid). In fact, it has been used in a parallel calculation carried 
out with REDUCE.
Here is the approximate group law:
\bea
x''^\alpha&=&x^\alpha+x'^\alpha+\eta_{\mu[\nu}{\delta_{\rho]}}^\alpha 
\;\epsilon'^{\nu\rho}x^\mu
+\eta_{\mu(\nu}{\delta_{\rho)}}^\alpha \;h'^{\nu\rho}
x^\mu+... \nn \\
\epsilon''^{\omega\rho}&=&\epsilon^{\omega\rho}+\epsilon'^{\omega\rho}-
\frac{1}{4}\eta_{[\alpha[\nu}{\delta_{\mu]}}^{[\omega}{\delta^{\rho]}}
_{\beta]}\;\epsilon'^{\mu\nu}\epsilon^{\alpha\beta}
-\frac{1}{4}\eta_{(\alpha(\nu}{\delta_{\mu)}}^{[\omega}{\delta^{\rho]}}
_{\beta)}\;h'^{\mu\nu}h^{\alpha\beta}+... \nn \\
h''^{\omega\rho}&=&h^{\omega\rho}+h'^{\omega\rho}-
\frac{1}{2}\eta_{(\alpha[\nu}{\delta_{\mu]}}^{(\omega}{\delta^{\rho)}}
_{\beta)}\;\epsilon'^{\mu\nu}h^{\alpha\beta}+... \label{grouplaw} \\
A''^\rho&=&A^\rho+A'^\rho+\left(\kappa c\;\;\eta_{(\alpha[\nu}{\delta_{\mu]}}^0{\delta
_{\beta)}}^\rho\right)\;\epsilon'^{\mu\nu}h^{\alpha\beta}+\eta_{\rho[\mu}{\delta_{\nu]}}^\alpha\; \epsilon'^{\mu\nu}A^\rho+ \nn \\
& &\frac{1}{2}\left( \kappa c\;\;\eta_{(\alpha(\nu}{\delta_{\mu)}}^{[0}
{\delta^{\rho]}}_{\beta)} \right)\; 
h'^{\mu\nu}h^{\alpha\beta}-\eta_{\rho(\mu}{\delta_{\nu)}}^\alpha\; h'^{\mu\nu}A^\rho+...\nn \\
\varphi''&=&\varphi'+\varphi-(m+\kappa q)c\;\;
\eta_{\mu[\nu}{\delta_{\rho]}}^0\; \epsilon'^{\mu\nu}x^\rho-mc\;
\eta_{\mu(\nu}{\delta_{\rho)}}^0 h'^{\nu\rho}x^\mu+
q\eta_{\nu\mu}A'^\nu x^\mu+ \nn \\
& &\frac{1}{2}\{\left(\frac{-1}{4}(m+\kappa q)c\;
\eta_{\rho[\sigma}{\delta_{\gamma]}}^0
\eta_{[\alpha[\nu}{\delta_{\mu]}}^{[\sigma}{\delta^{\gamma]}}_{\beta]}
-(m+\kappa q)c  \eta_{\sigma[\mu}{\delta_{\nu]}}^0
\eta_{\rho[\alpha}{\delta_{\beta]}}^\sigma) \right)\epsilon'^{\mu\nu}
\epsilon'^{\alpha\beta}x^\rho- \nn \\
& &mc \eta_{\sigma(\mu}{\delta_{\nu)}}^\sigma 
\eta_{\rho[\alpha}{\delta_{\beta]}}^\sigma h'^{\mu\nu}
\epsilon'^{\alpha\beta}x^\rho+q\;\eta_{\sigma\mu}\eta_{\rho[\alpha}{\delta_{\beta]}}^\sigma
A'^\mu \epsilon'^{\alpha\beta}x^\rho+ \nn \\
& &[\frac{-1}{4}(m+\kappa q)c\;\eta_{\rho[\sigma}{\delta_{\gamma]}}^0
\eta_{(\alpha(\nu}{\delta_{\mu)}}^{[\sigma}{\delta^{\gamma]}}_{\beta)}-
\frac{1}{2}\eta_{\rho\sigma}\left(-\kappa qc\;
\eta_{(\alpha(\nu}{\delta_{\mu)}}^{[0}{\delta^{\sigma]}}_{\beta)}\right)]
h'^{\mu\nu}h'^{\alpha\beta}x^\rho+ \nn \\
& &q\eta_{\sigma\mu}\eta_{\rho(\alpha}{\delta_{\beta)}}^\sigma
A'^\mu h'^{\alpha\beta}x^\rho\}+... \nn 
\eea

>From this law we can proceed following identical steps as in the pure
electromagnetic case and derive the approximate
quantization form $\Theta$, approximate Noether invariants,
Poincar\'e-Cartan form, Lagrangian, etc.. Let us write explicitly $\Theta$,
\bea
\Theta&=&d\varphi+ \nn \\
& &\{(m+\kappa q)c\;\eta_{\rho[\alpha}{\delta_{\beta]}}^0\;\epsilon^{\alpha\beta}+
mc\;\eta_{\rho(\alpha}{\delta_{\beta)}}^0
h^{\alpha\beta}-\nn \\
& &q\eta_{\alpha\rho}A^\alpha+ \frac{1}{4}(m+\kappa q)c
\left(\eta_{\sigma[\mu}{\delta_{\nu]}}^0\eta_{\rho[\beta}
{\delta_{\alpha]}}^{\sigma}+
\eta_{\sigma[\alpha}{\delta_{\beta]}}^0\eta_{\rho[\nu}
{\delta_{\mu]}}^{\sigma}\right)\epsilon^{\mu\nu}\epsilon^{\alpha\beta}- \nn \\
& &\frac{1}{4}mc \left(\eta_{\sigma(\mu}{\delta_{\nu)}}^0\eta_{\rho(\beta}
{\delta_{\alpha)}}^{\sigma}+
\eta_{\sigma(\alpha}{\delta_{\beta)}}^0\eta_{\rho(\nu}
{\delta_{\mu)}}^{\sigma}\right)h^{\mu\nu}h^{\alpha\beta}+ \nn \\
& &\frac{1}{2}[2(m+\kappa q)c\;\eta_{\rho(\mu}\eta_{\nu)[\alpha}
{\delta_{\beta]}}^0+mc\;\eta_{\rho[\alpha}\eta_{\beta](\mu}
{\delta_{\nu)}}^0]+\nn \\
& &\frac{q}{2}\eta_{\nu\rho}\eta_{\omega[\alpha}{\delta_{\beta]}}^\nu
\epsilon^{\alpha\beta}A^\omega-
\frac{q}{2}\eta_{\nu\rho}\eta_{\omega(\alpha}{\delta_{\beta)}}^\nu
h^{\alpha\beta}A^\omega
\;\}
dx^\rho+...\nn 
\eea

\noi and, before writing $d\Theta$, perform a change of variables in
order to take this presymplectic $2$-form to almost ``canonical" (or
standard) form \cite{footnote4}:

\bea
A^\alpha &\rightarrow& A^\alpha+\eta_{\sigma\gamma}(\epsilon^{\alpha\sigma}+
h^{\alpha\sigma})A^\gamma ... \nn\\
(m+\kappa q)c\; \epsilon^{0i}&\rightarrow&
(m+\kappa q)c\; \epsilon^{0i}+\frac{1}{2}(m+\kappa q)c\;(\epsilon^{ij}-\epsilon^{ji})\epsilon^{0j}+
g\delta_{ij}h^{00}\epsilon^{0j}-2(m+\kappa q)ch^{ij}\epsilon^{0j}+... \nn\\
h^{0j}&\rightarrow& h^{0j}+\frac{1}{2}(\epsilon^{ij}-\epsilon^{ji})h^{0i}\nn \\
h^{00}&\rightarrow& h^{00}-\frac{1}{4}\delta_{ij}h^{0i}\epsilon^{0j} \ \ ,
\eea  

\noi After this change $d\Theta$ acquires the expression:
\bea
d\Theta&=&d\,\epsilon^{0i}\wedge dx^0\left(-(m+\kappa q)c
(\epsilon^{0i}+h^{0i})\right)+ \nn \\
& &(m+\kappa q)c\; d\,\epsilon^{0i}\wedge dx^i +\nn \\
& &d\,h^{00}\wedge dx^0 2mc\left(1-2\,h^{00}\right) +\nn \\
& &d\,h^{0i}\wedge dx^0 \left(mc\;h^{0i}-(m+\kappa q)c\,\epsilon^{0i} 
\right)+\nn \\
& &d\,h^{00}\wedge dx^i mc\;h^{0i}+\nn \\
& &d\,h^{0j}\wedge dx^imc\left((-1+h^{00})\delta_{ij}-\,h^{ji}\right)
- \nn \\
& &mc\; d\,h^{ij}\wedge dx^i \;h^{0j} -\nn \\
& &q\;dA^0\wedge dx^0+\nn \\
& &q\;dA^i\wedge dx^i+...\nn 
\eea

The equations of motion can now be obtained \'a la Cartan by finding the
kernel of this Poincar\'e-Cartan-like form. 
Then, we have:
\bea
(m+\kappa q)c\frac{d^2\vec{x}}{dt^2}&=&q\left[\frac{d\vec{x}}{dt}\wedge\vec{\nabla}\wedge\vec{A}
-\partial_0\vec{A}-\vec{\nabla} A^0\right]+  \label{ecuElectrograv} \\
& &mc\left[\partial_0\vec{h}+\vec{\nabla}h^{00}
-\frac{d\vec{x}}{dt}\wedge\vec{\nabla}\wedge\vec{h}+ \right. \nn\\
& &\left.\frac{1}{4}\{-\partial_0(h^{00}\vec{h})+
\frac{d\vec{x}}{dt}\wedge\vec{\nabla}\wedge(h^{00}\vec{h})+
\partial_0(\vec{\vec{h}}\cdot\vec{h})-
\frac{d\vec{x}}{dt}\wedge\vec{\nabla}\wedge(\vec{\vec{h}}\cdot\vec{h})-\right.\nn\\
& &\left.2\vec{\nabla}({h^{00}}^2)+
\vec{\nabla}(\vec{h}\cdot\vec{h})\}+...\right]+ \nn\\
& &\frac{\kappa qc}{2}\left[\frac{1}{4}
\vec{\nabla}(\vec{h}\cdot\vec{h})+\partial_0\vec{h}
-\frac{d\vec{x}}{dt}\wedge\vec{\nabla}\wedge\vec{h}+...\right]\nn 
\eea

\noi The first line in (\ref{ecuElectrograv}) corresponds to the standard
(exact) motion of a  particle in the presence of an electromagnetic field, 
except for the value of the inertial mass, which is corrected 
by $\kappa q$. The second one reproduces the standard
gravito-electromagnetic force \cite{Wald}, i.e. the approximation in
which the gravitational field looks like an electromagnetic one. The third
and fourth are the first non-linear corrections to gravity. The fifth,
however, is quite new and represents a force that looks also like the
Lorentz force, at the present approximation, but generated by the
gravitational potentials, although proportional to $q$; it should not be
confused with the above-mentioned gravito-electromagnetic one. 
As far as the magnitude of the new Lie algebra co-homology constant
$\kappa$ is concerned, 
it is limited by experimental clearance for the difference between particle and 
anti-particle mass, which for the electron is about $10^{-8}m_e$. Even though this is a 
small value, extremely dense rotating bodies could be able to produce measurable forces. 
Conversely, a mixing of electromagnetism and gravity predicts a mass 
difference between charged particles and anti-particles, which could be experimentally 
tested by measuring, for instance, the Rydberg constant
($\sim \frac{m_{\rm antiproton} m_{\rm positron}}{m_{\rm antiproton} + 
m_{\rm positron}}$) through
the Lamb shift in anti-hydrogen \cite{Charlton}.

Let us remark the above-mentioned mentioned fact that one of the allowed
Lie-algebra co-homology extension parameters, that is $g$, has been
fixed to the particular value $g=mc$ in order to recover the standard
theory for $\kappa=0$. We might say that this requirement, along with
another condition of ``analyticity  in $q$" in the group law,
constitutes a group-theoretical setting of the (weak) {\it Equivalence Principle}.

Since the present theory has been formulated on the basis of our group
approach to quantization, the quantum version of it would proceed in a
rather straightforward manner.
We shall not insist any more on this particle
mechanical study while waiting for a wide generalization allowing for
field degrees of freedom. In fact, a 
natural yet highly non-elementary attempt at an extension of the present
theory to Quantum Field Theory 
is in course \cite{infinito}. A further generalization of the present
work in which the $U(1)$ subgroup of phase invariance is considered as a Cartan
subgroup of a larger internal symmetry group is also in order. Notice
that including the phase invariance in, for instance $SU(2)\otimes 
U(1)$, would result in additional phenomenology such as the production
of $Z_0$ particles out of gravity.    

\section*{Acknowledgments}
We desire to thank G. Gibbons for pointing out a relevant misprint in the statement of the
{\it no-go theorems} in a preliminary version of this letter. We also
thank C. Barcel\'o for valuable comments on the Equivalence Principle
and S. Vinitsky for exciting comments on the experimental test for the
new constant $\kappa$.


\begin{thebibliography}{99}
\bibitem{footnote1} The particular Hamiltonian vector field $X_H$ should
                    not be confused with $X^H$ defined previously
                    (\ref{XH}).
\bibitem{footnote2} We should remark that the symplectic correspondence
                    between functions and (Hamiltonian) vector fields can be kept at
                    the movement-space level provided that we restrict ourselves
                    to only those functions which are invariant under the action of $X^H$,
                    i.e. Noether invariants.
\bibitem{footnote4} This change might have been made in the group law
      by choicing a more appropriate Campbell-Hausdorff-like  formula but that
      choice is by no means a trivial task. We must stress on the fact that
      the quantum theory according to GAQ is free of these requirements as we
      associate quantum operators directly with group generators and the final
      output. i.e., the quantum representation is fully independent of any
      choice of co-ordinates in the group manifold. Even more, the classical
      theory can also be formulated in a fully invariant manner by using the
      Noether invariants as co-ordinates, but in this way the classical evolution
      is apparently hidden.
\end{thebibliography}
\end{document}